# Nanoparticle charge-transfer interactions induce surface dependent conformational changes in apolipoprotein biocorona

Achyut J Raghavendra[a], Nasser Alsaleh[b], Jared M. Brown[b*] and Ramakrishna Podila[a, c*].

Upon introduction into a biological system, engineered nanomaterials (ENMs) rapidly associate with a variety of biomolecules such as proteins and lipids to form a biocorona. The presence of 'biocorona' influences nano-bio interactions considerably, and could ultimately result in altered biological responses. Apolipoprotein A-I (ApoA-I), the major constituent of high-density lipoprotein (HDL), is one of the most prevalent proteins found in ENM-biocorona irrespective of ENM nature, size, and shape. Given the importance of ApoA-I in HDL and cholesterol transport, it is necessary to understand the mechanisms of ApoA-I adsorption and the associated structural changes for assessing consequences of ENM exposure. Here, we used a comprehensive array of microscopic and spectroscopic tools to elucidate the interactions between ApoA-I and 100 nm Ag nanoparticles (AgNPs) with four different surface functional groups. We found that the protein adsorption and secondary structural changes are highly dependent on the surface functionality. Our electrochemical studies provided new evidence for charge transfer interactions that influence ApoA-I unfolding. While the unfolding of ApoA-I on AgNPs did not significantly modify their cellular uptake and short-term cytotoxicity, we observed that it strongly altered the ability of only some AgNPs to generate reactive oxygen species in macrophages. Our results shed new light on the importance of surface functionality and charge transfer interactions in biocorona formation.

## Introduction

The sustainable implementation of engineered nanomaterials (ENMs) in biological applications, such as drug delivery and imaging, requires a comprehensive understanding of complex transformations and interactions at the nano-bio interface [1–4]. ENMs are known to rapidly associate with a variety of biomolecules (e.g., proteins, amino acids, and lipids) in any biological milieu to form a biocorona on their surface [5–12]. The presence of biocorona on ENMs imparts a new distinctive interactive surface, which ultimately determines the biological implications and fate of ENMs. Previous studies showed that ENM-biocorona (ENM-BC) leads to various physiological and pathological changes including protein aggregation, blood coagulation, and complement activation [5–12].

Although many physicochemical characteristics of ENMs have been reported to influence ENM-BC, it is not yet clear what combination of these properties could be used to predict ENM-BC formation and evolution [13, 14]. The composition of ENM-BC is known to strongly depend on ENM size and surface charge [5, 6, 9–11, 15–21]. Larger ENMs exhibit a preferential adsorption of higher molecular weight proteins whereas lower molecular weight proteins accumulate more on the surface of smaller ENMs. On the other hand, surface charge is expected to be a key factor in determining protein structural changes [9, 10, 12–15, 21]. The interactions between ENM surface and proteins could disrupt the structural integrity of proteins in ENM-BC and impact their function or elicit adverse immune responses. The surface coatings or functional groups are physically and chemically more active relative to the core of the ENM. For instance, citrate groups adsorbed on the surface of Ag nanoparticles (AgNPs) are negatively charged and provide electrostatic repulsion needed to prevent AgNP agglomeration. Any oppositely charged biomolecules (compared to the surface functional groups on the ENM) experience a natural electrostatic attraction to adsorb on ENM surface. Proteins with a stronger affinity to the ENM core can quickly displace the initially present functional groups and be irreversibly immobilized on the ENM surface by partial or complete denaturation [13, 15, 22]. A combination of many factors including van der Waals forces, hydrogen bridges, charge transfer and other hydrophobic interactions are known to drive protein denaturation [23–26]. Such factors naturally depend upon ENM physicochemical properties. It is imperative to deconvolute the influence of different physicochemical properties (e.g., surface charge and size) and understand how they differ from each other in influencing protein adsorption and denaturation in the ENM-BC.

Here, we experimentally investigated the influence of AgNP surface charges on adsorption and denaturation of apolipoprotein A-I (ApoA-I). Considering that ApoA-I is one of the most abundant proteins in AgNP-BC irrespective of AgNP size [20], its adsorption on AgNPs with different surface coatings could provide new insights into the role of surface charges in ENM-BC. ApoA-I is the major lipoprotein component of high-density lipoprotein (HDL). It adopts a shape similar to a horseshoe of dimensions 12.5x8x4 nm with high α-helix content [27–31]. The helices in ApoA-I are predicted to be amphipathic, with the hydrophobic (/hydrophilic) face mediating lipid (/aqueous) interactions. The thermodynamic drive to minimize

[a.] *Laboratory of Nano-biophysics, Department of Physics and Astronomy, Clemson University, Clemson, SC 29634, USA*
[b.] *Department of Pharmaceutical Sciences, Skaggs School of Pharmacy and Pharmaceutical Sciences, The University of Colorado Anschutz Medical Campus, Aurora, CO 80045, USA.*
[c.] *Clemson Nanomaterials Institute and COMSET, Clemson University, Clemson, SC 29634, USA.*
*Corresponding authors: rpodila@g.clemson.edu, jared.brown@ucdenver.edu

the aqueous exposure of the hydrophobic residues is one of the major factors in ApoA-I adsorption on AgNPs [32, 33]. We studied the interactions between ApoA-I and 100 nm AgNPs with four different coatings viz., citrate, polyvinylpyrrolidone (PVP), branched polyethylenimine (bPEI) and lipoic acid. These coatings were chosen to provide both negative (citrate, PVP, lipoic acid) and positive charged surfaces (bPEI) with different affinities for AgNPs. While lipoic acid interacts strongly through Ag-S bonds, other coatings (citrate, PVP, bPEI) are considerably weaker. Our light scattering and surface charge studies on 100 nm AgNPs revealed that the binding of ApoA-I on AgNPs is sensitive to their surface charge and functionality. While ApoA-I exhibited strong affinity for both positively and negatively charged AgNPs, its secondary structure exhibited more pronounced changes for two surface functionalities viz., lipoic acid and bPEI. In this article, we explain the observed secondary structural changes in terms of the electronic charge transfer, gleaned from electrochemical cyclic voltammetry experiments, between ApoA-I and functionalized AgNPs. Lastly, the displacement of positively charged bPEI by ApoA-I and the structural changes of ApoA-I on AgNP-lipoic acid were found to induce a significant increase in their ability to generate reactive oxygen species (ROS). Our results provide new insights into the role of AgNPs surface charge in ENM-BC formation and its influence on bio-response.

## Experimental methods

Aqueous solution of AgNPs (100nm, NanoXact, Nanocomposix) with citrate, polyvinylpyrrolidone (PVP), branched polyethylenimine (bPEI) and lipoic acid as capping agents were purchased. ApoA-I from human plasma (MW = 28.3KDa) was purchased from Sigma-Aldrich. AgNPs (50 μM) incubated overnight in different ApoA-I concentrations (0-4 g/L) were utilized to study hydrodynamic size and zeta potential using Zetasizer Nano ZS90 (Malvern Instruments). For transmission electron microscopy (TEM) characterization of the corona formation, AgNPs with ApoA-I biocorona were stained with 0.1% osmium tetroxide ($OsO_4$) by incubating for 30 minutes. These samples were then drop casted on 400 mesh Cu grid and TEM images were acquired using a Hitachi H-7600 microscope.

Electrochemical studies were performed using a Gamry reference 3000 electrochemical system. Cyclic voltammetry (CV) measurements were obtained in a three-electrode setup with Ag/AgCl as the reference electrode, platinum wire as the auxillary/counter electrode, and AgNPs as the working electrode. We used ApoA-I solution as the electrolyte to study charge transfer. It should be noted that we used concentration ranges based on the physiological levels of ApoA-I (~1-1.3 g/L) for electrochemical studies to avoid high current values that could result in artifacts in charge transfer measurements.

Circular Dichroism (CD) measurements were performed using Jasco spectropolarimeter (J-810) to analyze secondary structures of ApoA-I (300 μM) after the biocorona formation on AgNPs (50 μM). Samples were prepared for CD analysis with suspensions of AgNPs and ApoA-I corresponding to molar ratio of 1:600 and incubated at 40 °C for 8h. The CD spectra were measured at room temperature with wavelength range from 200-300nm for all the samples at a scan speed of 50nm/min. Background correction was applied with pure AgNP suspension.

Reactive oxygen species (ROS) were evaluated in RAW264.7 macrophages using dichlorofluorescein (DCF) via flow cytometry (by counting 10000 cells per sample using an Accuri C6 flow cytometer). Macrophages were grown to 90% confluence in RPMI 1640 medium and exposed to 30 μg/mL of AgNPs for 1h in serum free media to prevent any additional biocorona formation. All the measurements were performed in triplicates.

## Results and discussion

As shown in Fig. 1, TEM images confirmed that all four types of AgNPs (citrate, PVP, bPEI, and lipopic acid coated) are spherical in shape with a similar size ~100 nm in their dehydrated state. ApoA-I appeared to non-uniformly adsorb on to AgNPs (Figs. 1e-h) with strong preference for rough edges (see Figs. 1f and h and supplementary Figs. S1-S4). Given the small size of ApoA-I (hydrodynamic size ~ 8 nm) relative to 100 nm AgNPs, the biocorona possibly consisted of multiple ApoA-I layers. In some cases, ApoA-I was found to envelop AgNP aggregates in the dehydrated state under TEM similar to previous observations [37] (Figs. S1-S4).

Although all AgNPs have the same size in their dehydrated size (as evidenced by TEM), we found that their hydrodynamic sizes ($S_{HD}$) differed from each other in the following order: bPEI (~135 nm) > PVP (~120 nm)>lipoic acid (~108 nm)> citrate (~100 nm). This trend is expected based on the size of the surface coating molecules where polymers (bPEI and PVP) are significantly larger than the smaller citrate and lipoic acid groups. The difference in the surface coatings was also evident from their zeta potential (ζ) measurements where bPEI is positively charged with a relative low ζ ~8.89 mV while citrate (ζ~−39 mV), PVP (ζ~−37 mV), and lipoic acid (ζ~−28 mV) exhibited larger negative values. Based on the zeta potential measurements, negatively charged ApoA-I (ζ~−30–36 mV [37]) is expected to display strong electrostatic attraction towards bPEI coated AgNPs and possibly experience some repulsion from other negatively charged AgNPs.

We studied the evolution of $S_{HD}$ and ζ of AgNPs in the presence of ApoA-I (Figs. 2a and b). In the case of bPEI, ζ decreased rapidly and changed from a positive to negative value in the presence of ApoA-I. Indeed, ζ values (~-24 mV) at higher ApoA-I concentrations (> 2 g/L) suggest that ApoA-I stabilized AgNPs better than bPEI. A concomitant decrease in $S_{HD}$ of AgNP-bPEI suggested that ApoA-I has higher affinity for AgNP surface and ultimately replaced bPEI on the AgNP surface. The surface potential of AgNP-PVP decreased with increasing concentration of ApoA-1 similar to bPEI. However,

$S_{HD}$ showed a sharp increase with a rapid saturation at relatively low concentration of ApoA-I (<1 g/L). Unlike AgNP-bPEI and AgNP-PVP, a significant increase in $S_{HD}$ for AgNP-lipoic acid was observed only at higher ApoA-I concentrations (>2 g/L). The observed changes in both $\zeta$ and $S_{HD}$ were much slower for lipoic acid and can be rationalized in terms of relatively strong Ag-S interactions (binding energy ~217 kJ/mol [38]) between sulfur in lipoic acid and the surface of AgNPs. Interestingly, AgNP-citrate showed relatively smaller changes in $S_{HD}$ and $\zeta$ compared to other surface coatings. Nevertheless, in all cases, the increase in $S_{HD}$ did not indicate the formation of AgNP aggregates unlike in the dehydrated state (cf. Figs. S1-S4). We found that the changes in $S_{HD}$ and $\zeta$ saturated above 2 g/L for all surface coatings excepting bPEI and lipoic acid. While $S_{HD}$ for AgNP-lipoic acid appears saturated, its $\zeta$ did not suggest that protein adsorption was not saturated. When the average time between consecutive collisions ($\tau_c$) of ApoA-I with AgNPs is greater than ApoA-I configurational relaxation time ($\tau_R$), there is sufficient time for ApoA-I to unfold on the AgNP surface and thus could form a disorganized biocorona. This condition ($\tau_c > \tau_R$) occurs at low ApoA-I concentrations (<2 g/L) for AgNP-PVP and AgNP-citrate. Indeed, disorganized protein agglomerates can be clearly observed on AgNP-PVP surface (Fig. S2). However, at higher $\tau_c \sim \tau_R$, ApoA-I cannot completely unfold due to rapid collisions between ApoA-I and AgNPs. This results in a densely packed biocorona where ApoA-I retains much of its secondary structure. Based on the results described in Fig. 2, it may be expected that the biocorona layer present directly on the AgNP surface is densely packed at > 2g/L for all cases excepting bPEI and lipoic acid, and thus saturates resulting in further changes in $S_{HD}$ and $\zeta$. The collision frequency ($f_c = 1/\tau_c$) of ApoA-I molecules with AgNP surface may be calculated using the Smoluchowski equation [39]

$$f_c = 1/\tau_c = 2\pi DC d N_A \qquad (1)$$

where D= 120 $\mu m^2 s^{-1}$ is the room-temperature diffusion coefficient of the ApoA-I [40], d= 12.8 nm is the room-temperature hydrodynamic diameter of ApoA-I [41] and , $N_A = 6.023 \times 10^{23}$ $mol^{-1}$ is the Avogadro number, and C is ApoA-I concentration at which adsorption saturates. At C= 2 g/L, we find that $\tau_c \sim 90$ $\mu s$, which is on the same scale as protein unfolding timescales. It should be noted that the value of $\tau_c$ derived from the above analysis only provides an estimate for the unfolding timescales. Nevertheless, it could be used to infer that ApoA-I molecules will collide on a timescale of few tens of $\mu s$ at physiological concentrations ~1.3-1.5 g/L, leading to a rapid saturation in adsorption [42].

It is well known that some proteins change their conformation upon binding to nanoparticle surfaces. Previously, Cukalevski et al. studied the conformational changes of ApoA-I on polystyrene (PS NPs) and NIPAM/BAM (N-isopropylacrylamide-co-N-tert-butylacrylamide) small sized nanoparticles (<60 nm) with large curvature and different surface charges [33]. They found that uncoated and negatively charged PS NPs (diameter ~23-24 nm) slightly increased helical structure of ApoA-I in the range of 2-15 % whereas positively charged PS NPs (diameter ~57 nm) slightly reduced the amount of helical structure by ~10%. A similar study by Li et al. on 30 nm negatively charged AgNPs and ApoA-I [37] also found reduction of helical content by ~15-20%. Although these previous reports provide some preliminary understanding of ApoA-I interactions with ENMs, a controlled study with AgNPs of same size but different surface coatings is necessary to understand the influence of surface change. In this study, we used 100 nm AgNPs with different surface coatings to clearly distinguish the effects of surface charges and understand the mechanisms involved in ENM-induced protein unfolding. In our circular dichroism (CD) studies, we found that ApoA-I exhibited a significant decrease in the helical content by >40% on all 100 nm AgNPs. ApoA-I unfolding is more pronounced in our case due to the large AgNP size compared to previous studies [33, 37]. The CD spectra revealed that the changes in helical content were more prominent for AgNP-bPEI and AgNP-lipoic acid compared to PVP and citrate coatings. Indeed, the helical content of ApoA-I may be completely suppressed for these AgNPs (Fig. 3). These changes were accompanied by a concomitant increase in β-sheet and random chain structures in the following order: AgNP-PVP<AgNP-citrate<AgNP-lipoic acid~AgNP-bPEI.

As shown in Fig. 4, we performed cyclic voltammetry (CV) to develop a mechanistic understanding of surface charge dependent conformational changes observed in CD spectra (Fig. 3). To this end, AgNPs were immobilized on the glassy carbon working electrode and tested in ApoA-I solution as described in the experimental section. Proteins interact with ENM surfaces through intramolecular bonds, ionic bonds, and charge transfer [24, 26]. A stabilizing charge may be transferred between proteins and ENM surface depending upon their electronic energy levels, and the adsorbed proteins may undergo various conformational changes during the electron exchange process. Charge transfer processes and the relative differences between electronic energy levels of protein and ENM surface could be ascertained through peaks in current (i.e., charge flow) during a CV scan [18, 39]

In the CV scan of AgNPs without ApoA-I (dashed lines in Fig. 4a), we did not observe any peaks for all four surface coatings. The addition of ApoA-I to the electrolyte resulted in a change in voltammetric responses with the appearance of a new peak (blue arrows in Fig. 4a), which in turn increased the current considerably. This new peak cannot be attributed to desorption of adsorbed hydrogen (which is known to occur ~990 mV vs. Ag/AgCl electrode [39]) as the peak appeared at appreciably low potentials < 350 mV for all the coatings. We attribute this peak to stabilizing electron transfer between ApoA-I and AgNP surface. The electron transfer occurs only when the electronic energy levels of ApoA-I are in the vicinity of the AgNP's energy levels. The so-called Fermi energy ($E_F$) or the chemical potential of AgNPs (that serve as the working electrode in our CV measurements) is decreased (/increased) when the voltage in a CV scan is increased (/decreased). An electron transfer occurs when the energy levels of ApoA-I match the altered $E_F$ of AgNPs at a particular voltage in the CV scan. This peak voltage for electron transfer was found to be much lower for AgNP-bPEI and AgNP-lipoic acid (~60 mV) compared to AgNP-citrate and AgNP-PVP (~300 mV) suggesting that the surface coatings have considerable influence of electron transfer reactions between ApoA-I and AgNPs (Fig. 4a). The electrochemical charge transfer ($Q$) can be quantified by calculating the area enclosed by CV curves (Fig. 4a). We found that all the AgNPs showed an increase in $Q$ with increasing ApoA-I concentrations concurring with our hypothesis that the observed peak arises from electron transfer between ApoA-I and AgNPs. The normalized area charge density ($Q/cm^2$) at higher ApoA-I concentrations showed the following trend: AgNP-PVP>AgNP-citrate>AgNP-lipoic acid>AgNP-bPEI. This trend appeared to match the changes in CD spectra described earlier in Fig. 3.

An interesting feature in the CV scans was the presence of an irreversible charge transfer peak for AgNP-bPEI. For bPEI coating, a peak was observed only when the voltage was increasing in the CV scan indicative of irreversible charge transfer. On the other hand, CV scans for AgNP-citrate, AgNP-PVP, and AgNP-lipoic acid exhibited an observable valley (see red arrows in Fig. 4a) during the reverse voltage sweep suggesting that the observed $Q$ for these coatings does not entirely result from irreversible charge transfer reaction. This reversible peak may possibly be attributed to the formation of hydrogen bonds between ApoA-I and surface coatings (PVP, citrate, and lipoic acid). Given that bPEI showed completely irreversible stabilizing charge transfer, we expect greater conformational changes in ApoA-I. In other words, the loss of ApoA-I secondary structure on bPEI coated AgNPs precludes reversibility of charge transfer. Such an observation concurs with the CD spectra, which showed complete loss of $\alpha$–helical content in bPEI coatings. It could be expected that the disruption of stabilizing hydrophobic interactions in the interior of the protein results in an irreversible charge transfer and ensues in ApoA-I structural relaxation on AgNP-bPEI. On the other hand, AgNP-PVP and AgNP-citrate facilitate hydrogen bond formation with substantial retention of the helical content. Returning to Fig. 3, the loss in $\alpha$-helical content of ApoA-I on AgNP-lipoic acid cannot be explained based on charge transfer. It is possible that the ApoA-I molecule unfolds on AgNP-lipoic acid surface to increase its interaction with AgNP surface in order to break strong Ag-S interactions in AgNP-lipoic acid.

We assessed the ability of AgNPs with and without ApoA-I corona to generate intracellular ROS in RAW 264.7 macrophages after 1 h exposure as measured by DCF fluorescence. As shown in Fig. 5, we did not observe any significant changes (relative to control) in ROS generation for pristine AgNPs without ApoA-I biocorona irrespective of their surface coatings. However, we found that the addition of ApoA-I biocorona changed the response of AgNP-bPEI-ApoA-I relative to AgNP-bPEI without biocorona and AgNP-lipoic acid-ApoA-I relative to the control. Based on our CD and CV results, we attribute these changes to the adsorption and unfolding of ApoA-I on AgNP-bPEI and AgNP-lipoic acid. Additionally, AgNP-bPEI, ApoA-I biocorona resulted in an increased surface charge (accompanied by a change from positive to negative sign) and better stability (cf. Fig. 2b) that could have also been a contributing factor to the observed increase in ROS generation. Despite these changes in ROS generation, we did not observe any significant differences in the cytotoxicity of AgNPs before and after ApoA-I adsorption (Fig. S5). Furthermore, our flow cytometry data (Fig. S6) did not show significant changes for AgNPs upon the addition of ApoA-I biocorona. Based on these results (Figs. S5 and S6), changes in AgNP dissolution or uptake upon the addition of ApoA-I biocorona could be ruled out as possible cause for the observed differences in ROS generation.

## Conclusions

In summary, our dynamic and electrophoretic light scattering studies showed that ApoA-I displaces surface coatings such as citrate, PVP, and bPEI even at low concentrations (< 2 g/L). In case of AgNP-lipoic acid, strong Ag-S interactions inhibit ApoA-I adsorption for concentrations below 2 g/L. Circular dichroism studies showed a significant decrease in $\alpha$-helical content for all surface coatings with the complete disappearance of $\alpha$–helices for AgNP-bPEI and AgNP-lipoic acid. The changes in secondary structure concur with the observed charge transfer, measured using cyclic voltammetry, between ApoA-I and AgNPs. The unfolding of ApoA-I on AgNP-lipoic acid cannot be completely explained in terms of charge transfer. It is plausible that ApoA-I unfolds on the surface to lower its free energy and thereby break strong Ag-S interactions in AgNP-lipoic acid. Lastly, we found a significant increase in the ability of ApoA-I coated AgNP-bPEI and AgNP-lipoic acid to generate reactive oxygen species, which can be attributed to changes in surface charge and the unfolding of ApoA-I.

## Acknowledgements

R. P. and J. M. B. gratefully acknowledge support from NIH/NIEHS R03-ES023036 and R15-ES022766-01A1. R.P. is thankful to Clemson University for Start-Up funds.

**Figures and figure captions:**

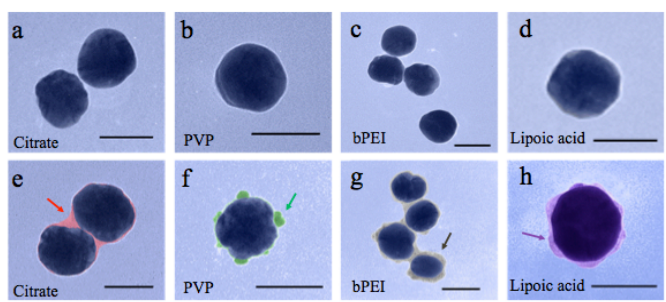

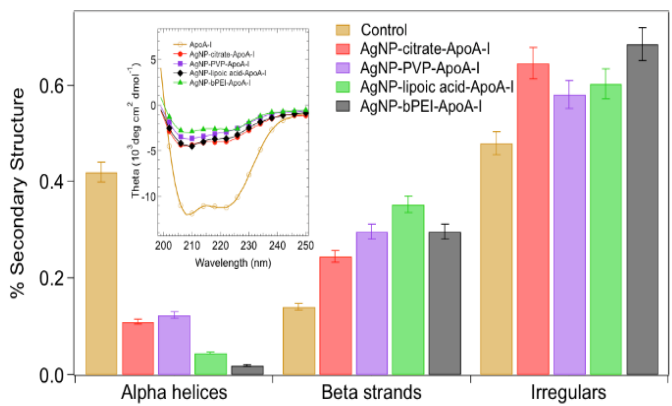

**Figure 1**: Transmission electron micrographs showing as-purchased AgNPs (a-d) with citrate, PVP, bPEI and lipoic acid coatings respectively. The presence of biocorona (see arrows in e-h) upon incubation with ApoA-I was assessed using $OsO_4$ staining as shown in (e-h). Scale bar is 100 nm.

**Figure 3**: Circular dichroism spectra for ApoA-I incubated with AgNPs (shown in the inset) of different surface coatings showed marked decrease in helical content with concomitant increase in beta sheets and irregular structures.

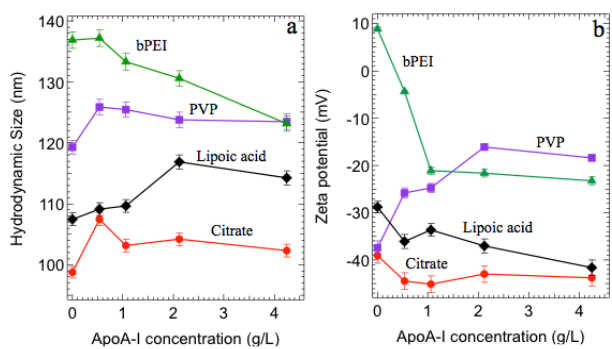

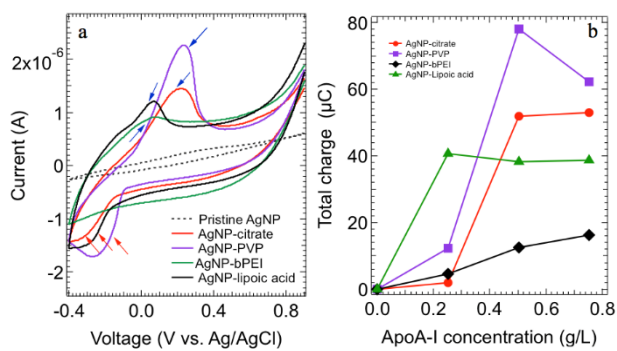

**Figure 2**: (a) The hydrodynamic size of AgNPs was found to change with increasing ApoA-I concentrations with saturation >2 g/L for AgNP-PVP, citrate, and lipoic acid coatings. AgNP-bPEI, however, did not show such saturation. (b) The zeta potential measurements showed clear changes indicating the displacement of surface coatings by ApoA-I. All surface coatings showed saturation in zeta potential changes >2 g/L excepting AgNP-lipoic acid. Based on (a) and (b), it could be inferred that AgNP-bPEI and AgNP-lipoic acid do not show saturation in protein adsorption even at high ApoA-I concentrations. It should be noted that the physiological concentration of ApoA-I is 1.3-1.5 g/L.

**Figure 4**: (a) Cyclic voltammetry scans showed the appearance of a peak (blue arrows) for AgNP electrodes in ApoA-I electrolyte suggesting the presence of charge-transfer stabilizing interactions. The peak for AgNP-bPEI occurred only in the forward scan (going from -0.4 to 0.8 V) indicating irreversible charge transfer. A valley (red arrows), representative of reversible charge transfer, was observed for other surface coatings on the reverse scan (0.8 to -0.4 V). (b) The total charge enclosed by the CV curves in (a) displayed clear increasing trends with ApoA-I concentration confirming that the charge-transfer occurs due to interactions between ApoA-I and AgNPs.

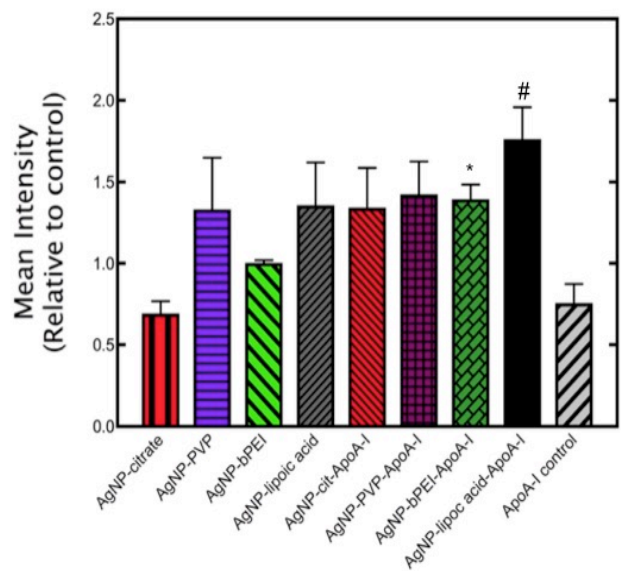

**Figure 5:** DCF assay was used to evaluate the ability of AgNPs with (+Apo) and without (-Apo) ApoA-I corona to generate reactive oxygen species (ROS). AgNP-bPEI-ApoA-I displayed significant increase in ROS generation compared to AgNP-bPEI while the response to AgNP-lipoic acid-ApoA-I was significantly higher than the control. These changes may be attributed the protein unfolding observed in AgNP-bPEI and AgNP-lipoic acid (cf. Fig. 3).

**Supplementary figures:**

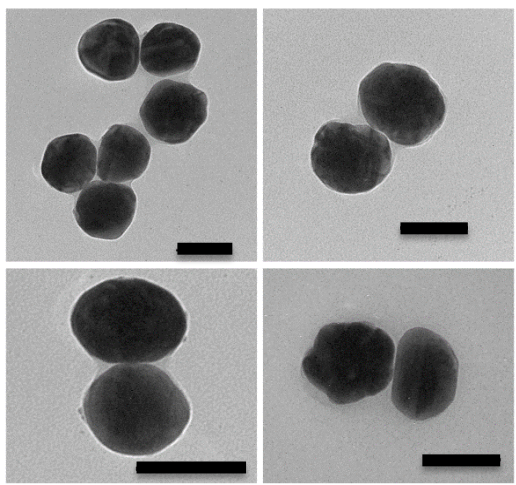

**Figure S1**: Transmission electron micrographs showing ApoA-I corona on AgNPs with citrate. Scale bar is 100 nm.

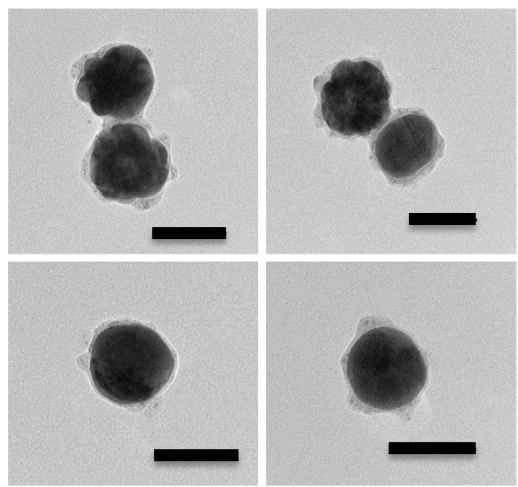

**Figure S3**: Transmission electron micrographs showing ApoA-I corona on AgNPs with branched polyethylenimine (bPEI). Scale bar is 100 nm.

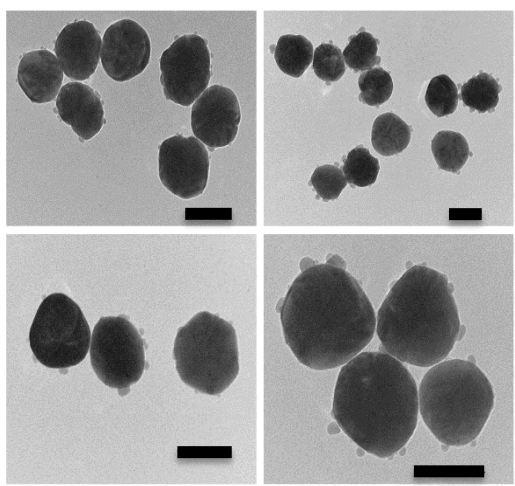

**Figure S2**: Transmission electron micrographs showing ApoA-I corona on AgNPs with polyvinylpyrrolidone (PVP). Scale bar is 100 nm.

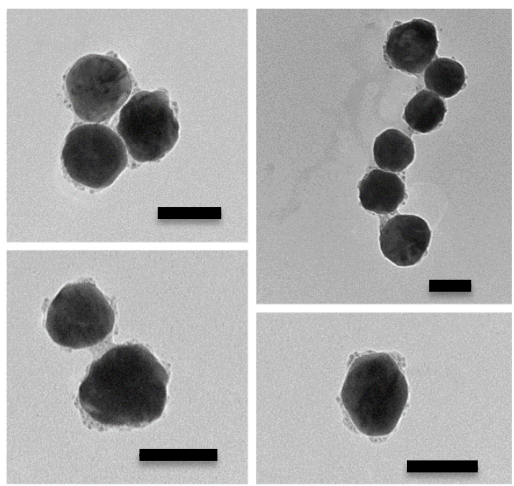

**Figure S4**: Transmission electron micrographs showing ApoA-I corona on AgNPs with lipoic acid. Scale bar is 100 nm.